\documentclass[preprint,12pt,titlepage,aps,showpacs]{revtex4}
\def\beq{\begin{equation}}
\def\eeq{\end{equation}}
\def\bea{\begin{eqnarray}}
\def\eea{\end{eqnarray}}

\def\be{\begin{equation}}
\def\ee{\end{equation}}
\usepackage{graphicx}

\begin{document}

\title{Radio Frequency Response of the Strongly Interacting Fermi Gases at Finite Temperatures}
\stepcounter{mpfootnote}
\author {S. Y. Chang}
\affiliation {Institute for Theoretical Physics,
		University of Innsbruck, Technikerstr. 25, A-6020 Innsbruck Austria}
\affiliation {Institute for Quantum Optics and Quantum Information of the Austrian
              Academy of Sciences,	ICT-Geb\"aude, Technikerstr. 21a ,
               A-6020 Innsbruck Austria}
\affiliation {Department of Physics and Institute for Nuclear Theory,
             Box 351560, University of Washington, Seattle, WA 98195 USA}
\date{\today}

\begin{abstract}
  The radio frequency spectrum of the fermions in the unitary limit at finite temperatures
is characterized by the sum rule relations. We consider a simple picture where the atoms are 
removed by radio frequency excitations from the strongly interacting states into a state of negligible interaction. 
We calculate the moments of the response function in the range of temperature $0.08 \epsilon_F < T < 0.8 \epsilon_F$ 
using auxiliary field Monte Carlo technique (AFMC) in which continuum auxiliary fields with a density dependent shift are used. 
We estimate the effects of superfluid pairing from the clock shift.
We find a qualitative agreement with the pairing gap - pseudogap transition behavior. 
We also find within the quasiparticle picture that in order for the gap to come into quantitative agreement with the previously
 known value at $T=0$ , the effective mass has to be $m^* \sim 1.43 m$. Finally, we discuss implications for the adiabatic sweep
of the resonant magnetic field.
\end{abstract}
\maketitle

 {\it Introduction:} Experiments with the dilute fermionic atomic gases such as those of Li$^6$ and K$^{40}$ atoms have seen  
great developments \cite{giorgini07}. Dilute Fermi gases provide a clean and controllable 
model for understanding different problems in a wide range of many body physics.
 Two species Fermi gas in the unitarity regime sits right in the middle of the BCS-BEC crossover regime
 and has been the focus of great interest. Radio frequency (RF) experiments can provide information on the
 interaction of the fermions by inducing excitations in the Fermi gas: in particular, the effects of fermionic superfluidity.
 Over the last few years, experimental measurements without the line broadening have become possible \cite{shin07}. 
However, in these early experiments the final state interaction effects seem non-negligible and they are rather hard to interpret for $T>T_c$:
in the experiment by Chin {\it et al.} \cite{shin07}, only the sharp free atom transitions are observed.
Recently, experimental data with negligible final state interactions and small
 collisional effects have been made available (see Ref. \cite{jin08,shirotzek08} for experiment and Ref. \cite{chen09} for theory). 
In particular, the double peak structure of the RF response \cite{shirotzek08} of the polarized Fermi gas
is qualitatively consistent with the phase separation. In this article, we study the RF response spectrum
and the effective mass relevant for these experiments by using an {\it ab initio} method.
 Here, we consider a situation where the atoms in the excited state remain non-interacting 
(this is the likely picture for K$^{40}$ around the Feshbach resonance at $202~G$ \cite{chen09}).
 This simplified situation allows us to calculate the clock shift and to estimate the effects of the energy excitation gap across the $T_c$.

One of the most puzzling aspects of the strongly interacting fermions is the temperature dependence of the quasiparticle gap $\Delta_{qp}$.
 Previous mean field analysis as well as the recent {\it ab initio} calculations \cite{stajic04,chen06,bulgac08,barnea08} suggest the existence 
of the so-called pseudogap above the $T_c$ (of which value at unitarity is most likely $\sim 0.15 \epsilon_F$ 
 \cite{prokofev08,bulgac08b}, $\epsilon_F$ = Fermi energy). The gap $\Delta_{qp}$  in the quasiparticle spectrum as a function of the $T$ was shown to have a slight dip around 
$T \sim T_c$ \cite{bulgac08,barnea08} and then to stay non-zero even at $T>T_c$ (at least up to $T \sim \epsilon_F$).
 This behavior is in sharp contrast to the condensate fraction calculated by two-body correlation 
$\Delta_0^2 \equiv \lim\limits_{r\rightarrow \infty} \frac{N}{2} \langle \int dr_1 \int dr_2  \Psi_1^\dagger(r_1 + r) \Psi_2^\dagger(r_2 + r) \Psi_2(r_2) \Psi_1(r_1) \rangle$
 that becomes non-zero only for $T < T_c$. Although the interpretation of $\Delta_{qp}$ as the order parameter does not apply for
all temperatures, $\Delta_{qp}$ is to show the transition from the superfluid pairing gap to the pseudogap or insulator gap for $T>T_c$. 
 We present in this article the temperature dependence of the RF spectrum in the unitarity regime in a simplified 
picture where the final state interactions can be ignored and the states remain sharply defined. 
We test the accuracy of the quasiparticle gap measurement from the RF response spectrum. Finally, we relate
the RF response to the adiabatic effect \cite{richardson97} at $T \approx T_c$ and discuss its experimental implications for 
 the observation of temperature changes during adiabatic sweep of the resonant field \cite{werner05,paiva09}.

 {\it Model:} The scenario we consider is the one in which the trapped atoms can occupy the Zeeman states labeled by $\sigma = 1,2,3$. 
Initially, only the states 1 and 2 are occupied by the atoms in equal number. The atoms in the states $1$ and $2$ are coupled
 with interaction given by the $s$-wave scattering length $a_s$. Our model is implemented in
 a cubic lattice box of coordinate points spaced by $\delta l$ in each direction. Thus, a cutoff in the momentum $k_c=\pi/\delta l$ is imposed
and the coupling strength is regulated by the relation: $1/g = m/(4\pi\hbar^2a_s) - 1/\Omega \sum_{|{\bf k}| =  0}^{k_c} 1/(2 \epsilon_{\bf k})$ \cite{prokofev08,bulgac08b}
 where $\epsilon_{\bf k} = \hbar^2 k^2/(2m)$ and $\Omega = $ volume of the system. 
In the absence of the RF perturbation, the state $3$ is not coupled to any of the other two states (unlike in the Ref. \cite{he09}). 
 This model is captured by the linear response theory of the unperturbed thermodynamic potential
\bea
{\cal H} & = &  \int d{\bf r}  \sum_{\sigma=1}^3 \Psi^\dagger_\sigma({\bf r})\left[-\frac{\hbar^2 \nabla^2}{2m} - \mu_\sigma\right] \Psi_\sigma({\bf r})  \nonumber \\
& + &  \int d{\bf r} g \Psi^\dagger_1({\bf r})\Psi_1({\bf r})\Psi^\dagger_2({\bf r})\Psi_2({\bf r}) \nonumber \\
& + &  \int d{\bf r} \sum_{\sigma=1}^3 \epsilon_\sigma \Psi^\dagger_\sigma({\bf r})  \Psi_\sigma({\bf r})~.
\label{Eqn_h}
\eea	
with the time dependent external perturbation ${\cal H}_{pert}(t)$.  Here, the $\Psi^\dagger_\sigma({\bf r})$
 and $\Psi_\sigma({\bf r})$ are the usual Fermi operators.   $\mu_\sigma$ are the chemical potentials corresponding to the
Zeeman levels with energies $\epsilon_\sigma$ ($\epsilon_3>\epsilon_2>\epsilon_1$).
When the state $2$ is coupled to the state $3$  by the RF perturbation, we assume a time dependent perturbation 
of the form ${\cal H}_{pert}(t) = {\cal Y} \cos(\omega t) e^{\eta t}$ with $\eta$ being a small and positive number
 and ${\cal Y} = i \int d{\bf r} [\Psi^\dagger_3({\bf r}) \Psi_2({\bf r}) - \Psi^\dagger_2({\bf r}) \Psi_3({\bf r})]$ is a hermitian operator. 
 The response function at any $T$ with respect to the perturbation operator ${\cal Y}$ is given by
\beq
S(\omega) = \frac{1}{\sum\limits_k e^{-E_k \beta}} \sum_{n,m}e^{-E_m \beta} \langle m| {\cal Y}^\dagger |n\rangle \langle n| {\cal Y} |m\rangle \delta(\omega - \omega_{nm})~.
\label{eqn_s}
\eeq
 where $|n\rangle$ and $E_n$ are the eigenstates and eigenvalues of the unperturbed Hamiltonian (Eq. \ref{Eqn_h}).
 And, we have $\beta \equiv 1/T$ and $\omega_{nm} \equiv (E_n-E_m)/\hbar $. Although methods to calculate directly the 
response function at zero temperature exist in the context of the Hartree-Fock theory and its density functional extension \cite{yabana96}, 
to obtain the response function in the full energy spectrum by an {\it ab initio} method remains as an open problem. 
Instead, we calculate the moments of the response function characterizing the energy transfer by the RF signal. 
 The moment of $i$-th power is defined as $M_i \equiv \hbar^{i+1} \int d\omega \omega^i S(\omega)$.
 $M_i$ can be evaluated by direct application of Eq. \ref{eqn_s}. The sum rules in compact notation are given as
 $M_0 =  \mbox{Tr}[ {\cal Y}^\dagger  {\cal Y} e^{-{\cal H}\beta}]/{\bf Z}$, $M_1 = \mbox{Tr}[ {\cal Y}^\dagger [{\cal H},{\cal Y}] e^{-{\cal H}\beta}]/{\bf Z}$,
 and $M_2 =  \mbox{Tr}[[{\cal Y}^\dagger,{\cal H}] [{\cal H},{\cal Y}] e^{-{\cal H}\beta}]/{\bf Z}$. 
Here, we have defined the partition function ${\bf Z} \equiv \mbox{Tr}[e^{-{\cal H}\beta}]$. 
These thermal averages are evaluated in the unperturbed basis where the system has unoccupied state $3$. 
Using the notation of the angled brackets as the thermal average, it can be shown that the zeroth moment 
$M_0 = \left\langle \int \Psi^\dagger_2({\bf r}) \Psi_2({\bf r}) d{\bf r} \right\rangle$. 
For the first (see also Ref. \cite{baym07,punk07}) and the second moments we have 
\bea
M_1 & = & -\left\langle\int d{\bf r} g{\cal N}_1({\bf r}) {\cal N}_2({\bf r}) \right\rangle + \delta \epsilon M_0  \\
 M_2 & = &  \left\langle \int d{\bf r} g^2{\cal N}_1^2({\bf r}) {\cal N}_2({\bf r}) \right\rangle +2\delta\epsilon M_1 - \delta\epsilon^2 M_0 ~.
\label{eqn_moments}
\eea
Here, ${\cal N}_\sigma({\bf r}) \equiv \Psi^\dagger_\sigma({\bf r})\Psi_\sigma({\bf r})$ and 
$\delta\epsilon \equiv (\mu_2 - \epsilon_2 -\mu_3 + \epsilon_3)$.  We restrict our study to the case where the chemical
 potentials measured with respect to the corresponding Zeeman levels are the same $\mu = \mu_\sigma - \epsilon_\sigma$ 
and we have $\delta \epsilon=0$. In this case, the atomic populations are balanced. 
The clock shift $\omega_s$ and the width $D$ of the response peak are obtained from 
\beq
\omega_s = \frac{M_1}{M_0}~~~ ,~~~ D = \sqrt{\frac{M_2}{M_0}-\left[\frac{M_1}{M_0}\right]^2}
\eeq
Thus, $\omega_s(a_s = \infty)$ is the same as the difference of the clock shifts 
$\delta\omega_s = \frac{1}{M_0}(M_1(a_s = \infty) - M_1(a_s = 0))$.
$M_0$, $\omega_s$ and $D$ can be interpreted as the parameters of the Gaussian fit \cite{paris02} to the response function:  
$S_G(\omega) = M_0/\sqrt{2\pi D^2} e^{-(\omega -\omega_s)^2/(2 D^2)}$.  From the experiments \cite{shin07}, 
this fit seems to be sufficiently accurate assuming that there are no impurities such as free atoms. 
For this case, $M_1$ and $M_2$ are positive for all $T$. When $g=0$ (or $1/a_s = -\infty$), the response peak has zero shift $\omega_s = 0$ and
the width  $D =0$. $\omega_s$ measures the energy transfer per particle
in the coherent rotation of the initial $|12\rangle$ state into the $|1\beta\rangle$ state with $|\beta\rangle$ being a point in the
pseudospin Bloch sphere. However, since $|\beta \rangle$ is not an eigenstate of the Hamiltonian the coherence is lost
due to the interactions with subsequent increasing of $D$ \cite{baym07}. Thus we have non-zero $\omega_s$ and $D>0$ for the
interacting system ($g = g_{12} \ne0$) even when the level couplings $g_{13} = g_{23} = 0$. 

{\it Method:} For the lattice model, it can be shown that around the resonance ($a_s \approx \pm\infty$) 
the sign problem can be avoided with the finite and negative regulated coupling $g$. 
 The complex phase problem returns when the molecule size ($\sim a_s$) in the BEC regime becomes smaller than the lattice spacing $\delta l$ and 
the coupling constant $g$ becomes positive.  For the finite $T$ calculations we use the grand canonical formalism \cite{hirsch83,bulgac06}.
Here, the thermal operator $e^{-{\cal H}\beta}$ can be applied in the single particle basis after Hubbard-Stratonovich (HS) transformation.
 The resulting multidimensional integration of the auxiliary variables is evaluated by the Monte Carlo methods.
 Typically, the inverse temperature $\beta$ is sliced into a few hundreds of smaller steps $\delta\beta$.
\begin{widetext}
\beq
e^{- g \int \Psi^\dagger_1({\bf r})\Psi_1({\bf r})\Psi^\dagger_2({\bf r})\Psi_2({\bf r}) d{\bf r}\delta \beta} 
=  \prod\limits_{\bf r} \int\limits_{-\infty}^{\infty} dx({\bf r}) \frac{e^{-x({\bf r})^2/2}}{\sqrt{2\pi}}
e^{x_0 x({\bf r}) - x_0^2/2} e^{-[(x({\bf r})-x_0)\sqrt{-g\delta\beta}+\frac{g \delta\beta}{2} ] (\Psi^\dagger_1({\bf r})\Psi_1({\bf r})+\Psi^\dagger_2({\bf r})\Psi_2({\bf r}))}~.
\label{eqn_hs}
\eeq
\end{widetext}
 The thermalization of the stochastic samples can be optimized by shifting the center of the auxiliary fields. 
The mean shift of the auxiliary fields can be derived from the minimal condition of the weight function.
 Thus the shift of the fields becomes $x_0 = \sqrt{-g\delta\beta} n$ \cite{stoicheva07} with $n = k_F^3/(3\pi^2)$. 
In the practice, some freedom is given to the choice of $x_0$ within a factor of order one. This shift of the auxiliary field modifies 
the HS transformation as shown in the Eq. \ref{eqn_hs}. The Monte Carlo integration with the probability density given by Eq. \ref{eqn_hs} is carried out in the following steps: 
Firstly, the value of the field $x({\bf r})$ is tentatively given by direct Gaussian sampling.
 Secondly, the operator part $e^{-[(x({\bf r})-x_0)  \sqrt{-g\delta\beta} + g\delta\beta/2] (\Psi^\dagger_1({\bf r})\Psi_1({\bf r})+\Psi^\dagger_2({\bf r})\Psi_2({\bf r}))}$
 is expressed in the single particle orbitals.
 Finally, the resulting probability density associated with the propagator is evaluated as the determinant of a matrix \cite{hirsch83},
 while the factor $e^{x_0 x({\bf r}) - x_0^2/2}$ contributes as a simple numerical factor to the total probability density. 
This configuration is accepted or rejected by the Metropolis algorithm. In order to obtain the acceptance probability of $30 \sim 70 \%$,
 only a small fraction of the field variables are updated in each trial. 
At $T \sim 0.1 \epsilon_F$ just $0.5\%$ of the field variables are updated while at $T \sim 0.4 \epsilon_F$, $\sim 3\%$ of the auxiliary variables can be
 updated in each step.  The benefit of shifting the center of the auxiliary fields is that the samples start out closer to the thermalized
 configuration avoiding long initial thermalization runs. However, this shift has negligible effects in reducing the statistical 
fluctuations of the thermalized samples. Nor does this shift appear to ameliorate the sign problem of the unbalanced system with $\mu_1 \ne \mu_2$. 
The Monte Carlo simulations produce as output the one-body density matrix $n_\sigma({\bf k} ,{\bf k}',\{x({\bf r})\})$ according to
the sampled configuration $\{x({\bf r})\}$. Through Fourier transform, it can be connected to other quantities in the coordinate space.

\begin{figure}
\includegraphics[angle=0,width= 8cm]{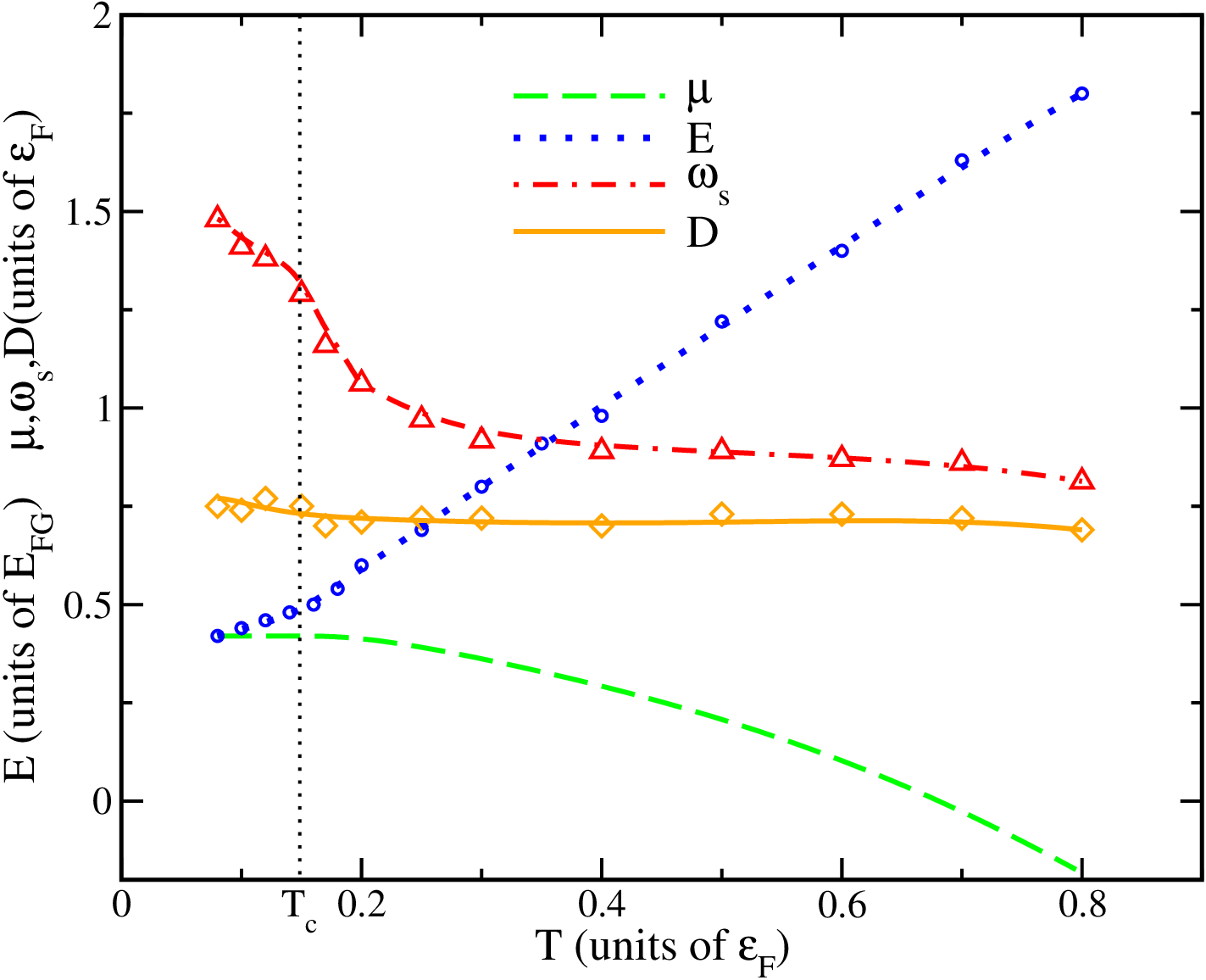}
\caption{(Color online) Here the energy per particle $E$ in the units of $E_{FG} = 0.6 \epsilon_F$ is shown as circles (dotted line).
 All the other quantities are in the units of $\epsilon_F$. The dashed line corresponds to the chemical potential $\mu$.
Both $E$ and $\mu$ are in good agreement with the known results \cite{bulgac06}.
The triangles(dot dashed line) represent the RF clock shift $\omega_s(a_s = \infty)$ and the diamonds (line) the width $D$.  
$\omega_s$ has a `bump' at $T \lesssim 0.2 \epsilon_F$ and
then remains more or less constant for higher temperatures. 
The width $D$ of the response peak is almost featureless. It remains more or less constant $D \sim 0.7 \epsilon_F$ at all $T$.
These results were obtained for $7^3$ and $9^3$ lattice boxes with the
 filling density of $\sim 0.09$. The smooth fitting curves were generated by Bezier algorithm.  
The size of the symbols correspond approximately to the statistical errors. The vertical dotted line corresponds to the 
probable position of $T_c$ \cite{prokofev08,bulgac08b}. } 
\label{fig_sum}
\end{figure} 

\begin{figure}
\includegraphics[angle=0,width= 6cm]{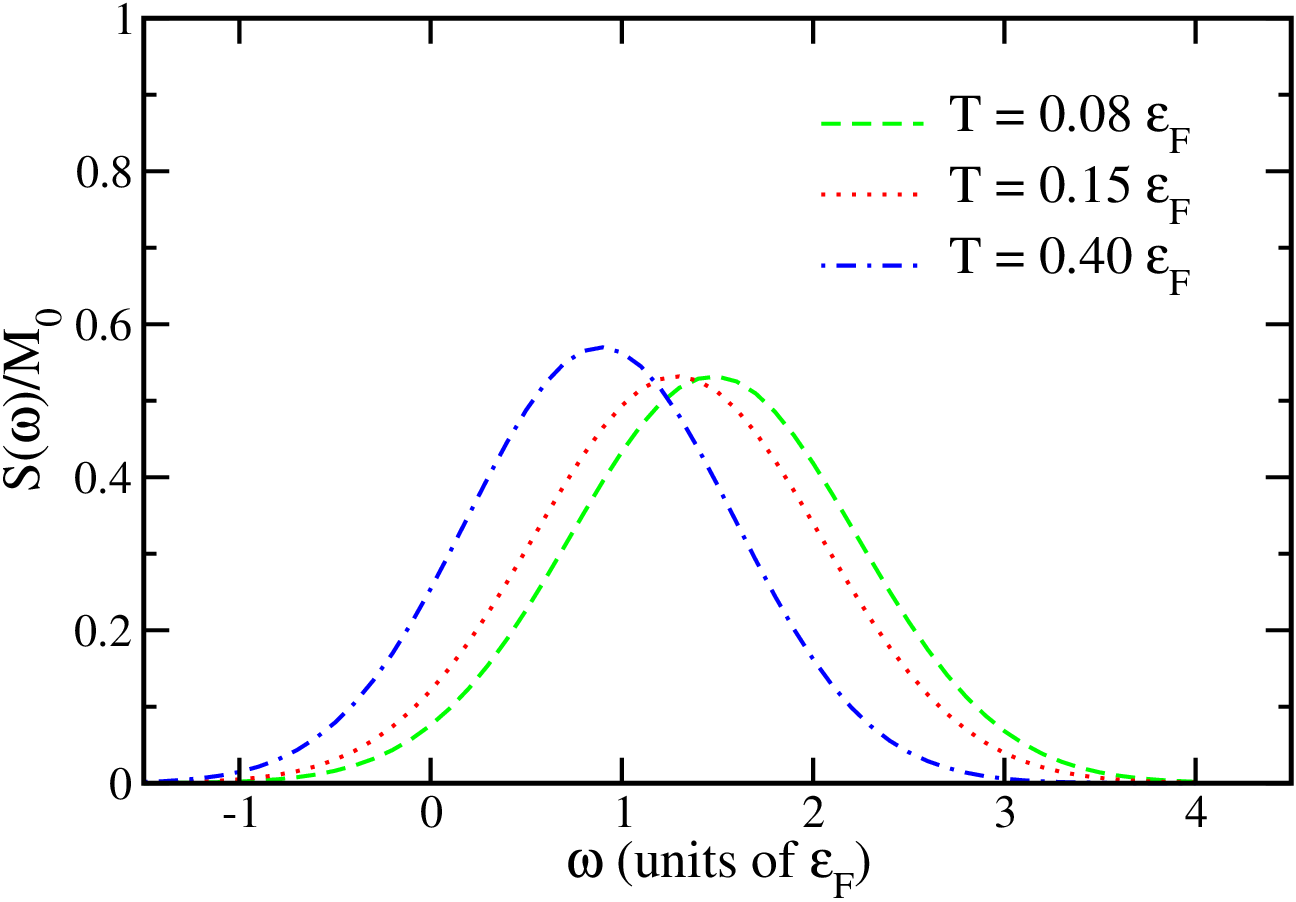}
\caption{(Color online) Temperature dependence of the RF response $S(\omega)$ (by Gaussian fit described in the text).} 
\label{fig_s}
\end{figure} 

\begin{figure}
\includegraphics[angle=0,width= 6cm]{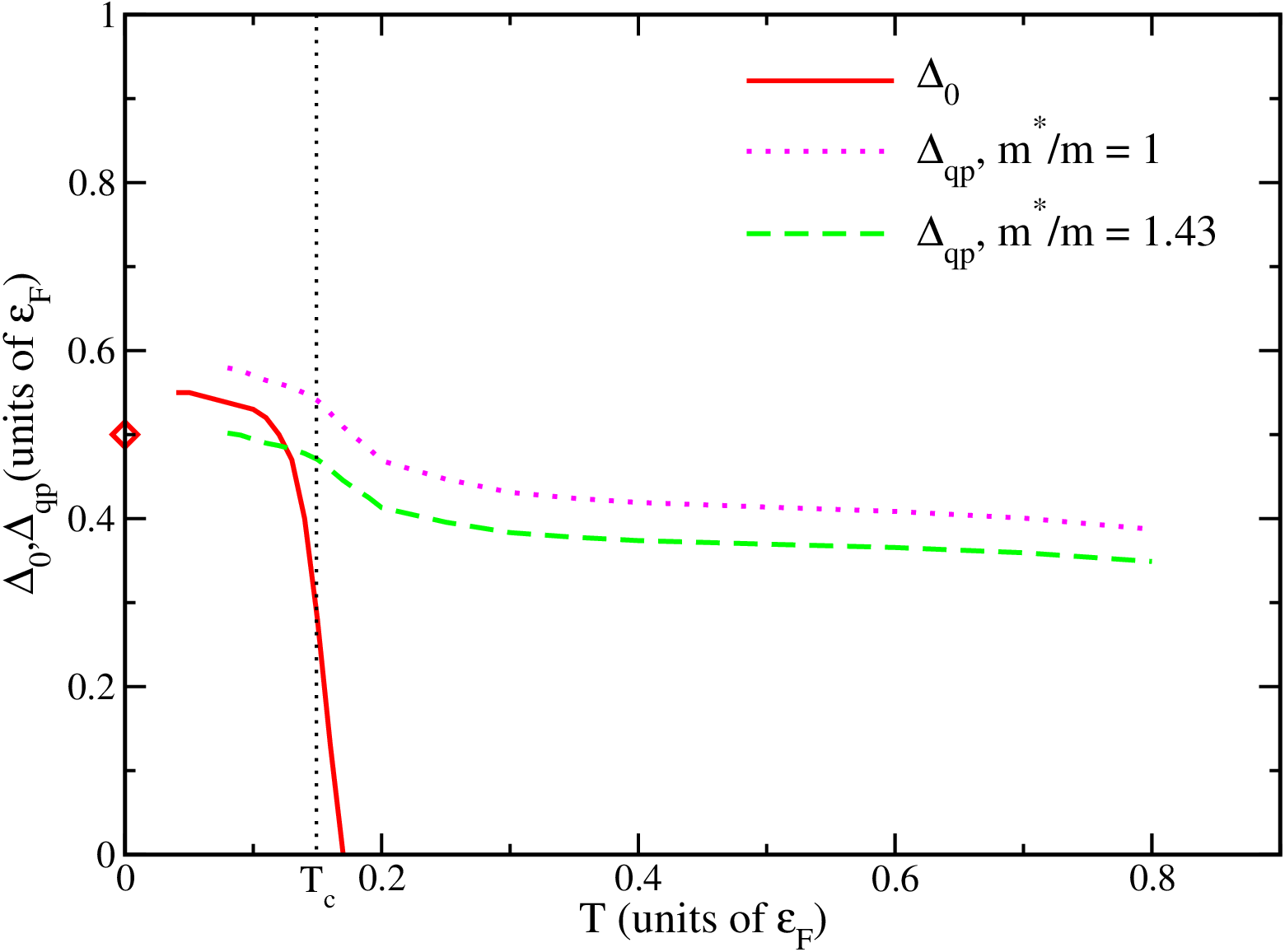}
\caption{(Color online) Here the pairing gap $\Delta_{qp}$ calculated using the Eq. \ref{eqn_mf} is shown for different
effective masses: dotted line for $m^*/m = 1$ and dashed line for $m^*/m = 1.43$.  
We notice that $\Delta_{qp}$ has a bulge at $T \lesssim T_c$ matching qualitatively with the behavior of the order parameter $\Delta_0$
(line from Ref. \cite{barnea08} where dynamic mean field method was used). Then for $T > 0.2 \epsilon_F$, $\Delta_{qp}$
 smoothly decays showing the transition into the gap insulator phase. $\Delta_{qp} = 0.50(2) \epsilon_F$ \cite{carlson05} at $T = 0$ is also shown (diamond) for the comparison purpose.} 
\label{fig_gap}
\end{figure} 

\begin{figure}
\includegraphics[angle=0,width= 8cm]{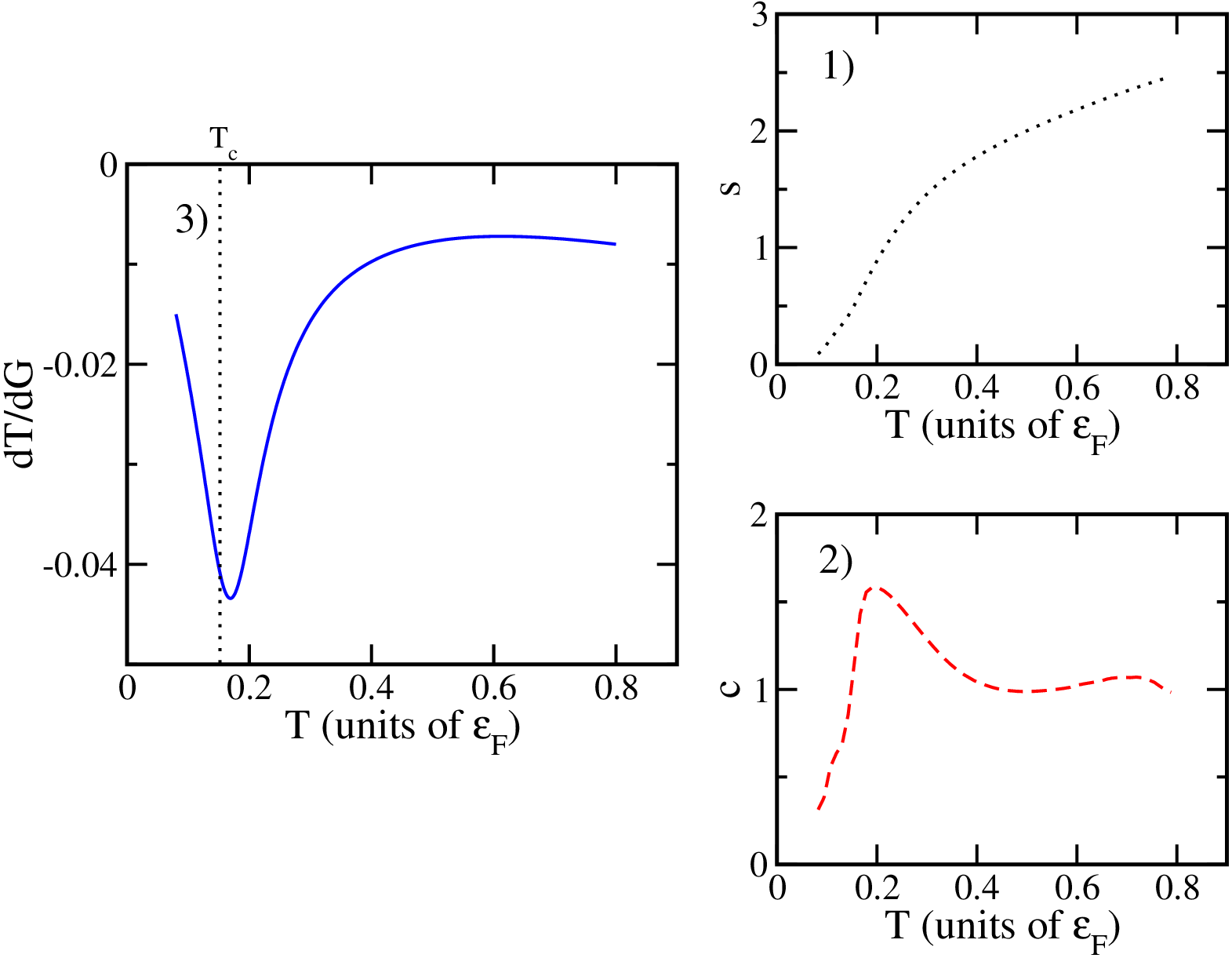}
\caption{(Color online) In the plot 1), the entropy per particle $s$ is shown using the method 
described in the Ref. \cite{bulgac08b}. The plot 2) shows the corresponding heat capacity calculated from
$c = T\frac{\partial s}{\partial T}$. The adiabatic behavior of the temperature for small changes of the coupling $G=g/\delta l^3$ is
shown in the plot 3). The quantities shown here are in dimensionless units. } 
\label{fig_cool}
\end{figure}

{\it Results and Discussion:} In the simple BCS mean-field picture\cite{yu06,zoller00,leskinen08}, the frequency shift is 
\beq
\omega_{MF}  =   -g' \frac{n}{2} - \frac{2 \Delta_{qp}^2}{g'n}~.
\label{eqn_mf}
\eeq
where $ g' = \frac{m}{m^*} g$, $n$ is the particle density and $m^*$ is the effective mass. 
In principle, $g'$ can have temperature dependence through the effective 
mass $m^*(T)$. In Ref. \cite{bulgac07} the calculated effective mass stays more or less close to $\sim m$ in 
the range of temperature $0.1 \epsilon_F \lesssim T \lesssim 0.9 \epsilon_F$ \cite{bulgac08}. In our case,
we need the ratio $m^*/m \sim 1.43$ in order to extrapolate to the known $T=0$ result for the quasiparticle gap.
Then we keep the same effective mass for $T>0$. 
Our results are summarized in the figures \ref{fig_sum},\ref{fig_s}, \ref{fig_gap}, and \ref{fig_cool}. 
These results are obtained for the lattice box of volumes $7^3$ 
and $9^3$ with the periodic boundary conditions. Here, the finite size dependence is within the error bars.
  $\Delta_{qp}$ shown in Fig. \ref{fig_gap} is obtained from fitting $\omega_s$ to the Eq. \ref{eqn_mf}.
Even without considering the mass renormalization ($m^*/m = 1$), $\Delta_{qp}$ at $T\lesssim T_c$ has a qualitative tendency close to the known
value of the $\Delta_{qp}$ at zero temperature \cite{carlson05}. In order to optimize the fit, we need to adjust the mass
by $m^*/m \sim 1.43$ as earlier mentioned. We notice that the $\Delta_{qp}$ has a bulging feature at low temperature
(Pomeranchuk effect as discussed later) and then decays slowly for $T> 0.2\epsilon_F$. This is somewhat similar to the pseudogap behavior discussed in Ref. \cite{stajic04,chen06}: 
$\Delta_{qp}^2 = \Delta_{sc}^2+\Delta_{pg}^2$ where $\Delta_{sc}=\Delta_0$ is the
 superconducting order parameter with non-zero values at $T\le T_c$ and $\Delta_{pg}$ is the pseudogap which
is non-zero for $T\le T^*$ with $T^* > T_c$. However, in our case there is no clear evidence of $T^*$. In fact,
at $T > T_c$ we enter into the regime of the gap insulator phase as discussed in Ref. \cite{barnea08}.
 The width of the response $D$ shows the broadening effect due to the decoherence
as discussed earlier (Fig. \ref{fig_sum},\ref{fig_s}). $D$ remains more or less constant in the range of the temperature that we considered. 
In the experiment of reference \cite{jin08} the population density of the K$^{40}$ atoms in the state 3 is measured
 as a function of the energy $\omega$ and the momentum $k$ of the ejected particles. In this case, the momentum contribution by
 the RF signal is considered negligible. Thus the measured momentum is a good estimate of that of the non-perturbed
 system. Here, the measured $\Delta_{qp}/\mu  \approx 0.75$ at $T \approx T_c$ while from our estimates
this ratio is $\sim 1$. In this experiment, however, the trap is spatially inhomogeneous causing the
$k_F$ to vary locally.  In another experiment (see Ref. \cite{shirotzek08}), by bimodal spectral response,
the ratio $\Delta_{qp}/\mu  \approx 1$ has been measured at $T \lesssim T_c$ in closer agreement with 
our theoretical results.

 In comparison to the earlier theoretical works without the final state interaction effect (Ref. \cite{stajic04, kinnunen04,stoof08}), 
both the RF clock shift and the width of the response function are found to be larger. 
Also, in our case $\omega_s$ does not approach zero value in the studied temperature range.
Unlike in the two channel models \cite{stajic04,he09,kinnunen04,he05}, we omitted the free atom contributions
 that produce a sharp peak at the zero frequency shift.

 Connection to the Pomeranchuk effect for the lattice fermions was made in the Ref. \cite{werner05,paiva09}.
Here, anomalous double occupancy of the repulsively interacting fermions in the two dimensional lattice at half filling
 was discussed. Analogously, for the unitary limit Fermi gas we can relate the double occupancy to the RF clock
shift by $M_0 \frac{\omega_s}{-g}$. From the Fig. \ref{fig_sum}, we notice an enhancement of the density overlap (double occupancy)
at $T \lesssim 0.2 \epsilon_F$. In this case, the adiabatic behavior of the temperature for small changes of the coupling $g$
can be obtained from $\frac{\partial T}{\partial g} = \frac{T}{-2 g c(T)} \frac{\partial \omega_s(T) }{\partial T} $.
The change in the coupling constant is $\delta g > 0$ (while $g < 0$) when the unitary limit is crossed from the BCS to the
 BEC side and an enhancement of the adiabatic effect that lowers the temperature occurs at $T \gtrsim T_c$ (Fig. \ref{fig_cool}).
This is qualitatively different from the \cite{bulgac08,barnea08} where the non monotonic $\omega_s(T)$ can produce temperature increase. 
Thus, precise measurement of the temperature at around $T \approx T_c$ with adiabatic changes of the
resonant magnetic field can lead to the verification of the temperature dependence of $\omega_s(T)$ (and also that of $\Delta_{pg}(T)$). 
Since there is no transfer of heat during this adiabatic process, this is not a cooling nor a heating effect.
Similar treatment within the mean field formalism is also discussed in the Ref. \cite{chen05}.

 {\it Concluding Remarks:} In summary, we have studied the RF response function of the unitarity Fermi gas in a fully three dimensional system by
 a numerical method free of any uncontrolled approximation. We found an unambiguous
 signature of the pairing gap at low temperatures and also the pseudogap behavior at $T\gtrsim T_c$
that seems to differ qualitatively from some of the existing works. We described the adiabatic temperature effect which
could lead to the experimental confirmation of the pseudogap features.
We also found that within the quasiparticle picture a rather heavy effective mass has to be assumed.
For the calculations, we relied on the grand canonical formalism with small statistical errors. There are finite temperature canonical formalisms \cite{ormand94} where,
 at least in principle, the usual odd-even staggering of the energy as a function of the particle numbers 
can be used to extract the pairing gap. However, the errors are known to be larger and the computational demands 
much higher. We have shown that the RF spectroscopy is a useful way to observe the pseudogap behavior.
 I am grateful to  M.M. Forbes, A. Bulgac, N. Barnea, W. Yi, and N. Trivedi for useful comments and discussions. This work 
was  supported by the U.S. Department of Energy under Grants DE-FG02-00ER41132 and DE-FC02-07ER41457
and DARPA grant BAA 06-19.


\begin{thebibliography}{}
\expandafter\ifx\csname natexlab\endcsname\relax\def\natexlab#1{#1}\fi
\expandafter\ifx\csname bibnamefont\endcsname\relax
  \def\bibnamefont#1{#1}\fi
\expandafter\ifx\csname bibfnamefont\endcsname\relax
  \def\bibfnamefont#1{#1}\fi
\expandafter\ifx\csname citenamefont\endcsname\relax
  \def\citenamefont#1{#1}\fi
\expandafter\ifx\csname url\endcsname\relax
  \def\url#1{\texttt{#1}}\fi
\expandafter\ifx\csname urlprefix\endcsname\relax\def\urlprefix{URL }\fi
\providecommand{\bibinfo}[2]{#2}
\providecommand{\eprint}[2][]{\url{#2}}

\bibitem{giorgini07} See for example summaries by S.~Giorgini, L.~P.~Pitaevskii and S.~Stringari, Rev. Mod.
Phys. {\bf 80}, 1215 (2007), and by R.~Grimm  cond-mat/0703091 (2007). 
\bibitem{shin07}  Y. Shin, C. H. Schunck, A. Schirotzek, and W. Ketterle, Phys. Rev. Lett. {\bf 99}, 090403 (2007). S. Gupta, Z. Hadzibabic, M. W. Zwierlein, C. A. Stan, K. Dieckmann, C. H. Schunck, E. G. M. van Kempen, B. J. Verhaar,  W. Ketterle, Science {\bf 300} 1723 (2003).  M. Greiner, C. A. Regal, and D. S. Jin, Phys. Rev. Lett. {\bf 94} 070403 (2005).
\bibitem{jin08} J. T. Stewart, J. P. Gaebler, and D. S. Jin, Nature {\bf 454}, 744 (2008).
\bibitem{shirotzek08} A.~Schirotzek, Y.~I.~Shin, C.~H.~Schunck, and Wolfgang Ketterle, Phys. Rev. Lett. {\bf 101}, 140403 (2008).
\bibitem{chen09} Q.~Chen, and K. Levin, Phys. Rev. Lett. {\bf 102}, 190402 (2009).
\bibitem{stajic04} J.~Stajic, J.~N.~Milstein, Q.~Chen, M.~L. Chiofalo, M.~J. Holland, and K.~ Levin, Phys. Rev. A {\bf 69}, 063610 (2004).
\bibitem{chen06} Q.~Chen, I.~Kosztin, B. Jank\'o, and K.~Levin, Phys. Rev. B {\bf 59}, 7083 (1999). 
A.~Perali, P.~Pieri, G.~C. Strinati, and C. Castellani, Phys. Rev. B {\bf 66}, 024510 (2002).
Q.~Chen, J.~Stajic and L.~Levin, Low Temp. Phys. {\bf 32(4)}, 406-423 (2006).
\bibitem{bulgac08} A. Bulgac {\it et al.}, arXiv:0801.1504v1 (2008).
\bibitem{barnea08} N. Barnea, Phys. Rev. A {\bf 78}, 053629 (2008). 
\bibitem{prokofev08} E. Burovski, N. Prokof'ev, B. Svistunov, and M. Troyer, Phys. Rev. Lett. {\bf 96},
 160402(2006). E. Burovski,E. Kozik, N. Prokof'ev, B. Svistunov, and M. Troyer, Phys. Rev. Lett. {\bf 101}, 090402 (2008).
\bibitem{bulgac08b} A. Bulgac, J. E. Drut and P. Magierski, Phys. Rev. A {\bf 78}, 023625 (2008).
\bibitem{richardson97} R. C. Richardson, Rev. Mod. Phys. {\bf 69}, 683 (1997).
\bibitem{werner05} F. Werner, O. Parcollet, A. Georges, and S.~R. Hassan, Phys. Rev. Lett. {\bf 95} 056401 (2005).
\bibitem{paiva09} T. Paiva, R. Scalettar, M. Randeria, and N. Trivedi, arXiv:0906.2141v1 (2009).
\bibitem{he09} Y.~He, C.-C.~Chien, Q.~Chen, and K~. Levin, Phys. Rev. Lett. {\bf 102} 020402 (2009).
\bibitem{yabana96} K. Yabana and G. F. Bertsch, Phys. Rev. B {\bf 54}, 4484 (1996).
K. Yabana and G.F. Bertsch, Phys. Rev. A {\bf 60}, 1271 (1999).
\bibitem{baym07} G. Baym et al. Phys. Rev. Lett. {\bf 99}, 190407 (2007).
\bibitem{punk07} M. Punk and W. Zwerger, Phys. Rev. Lett. {\bf 99}, 170404 (2007).
\bibitem{paris02} M.~W.~Paris and V.~R.~Pandharipande, Phys. Rev. C {\bf 65}, 035203 (2002).
\bibitem{hirsch83} J. E.~Hirsch, Phys. Rev. B {\bf 28}, 4059(R) (1983).
G. H.~Lang, C. W. Johnson, S. E. Koonin and W. E. Ormand,  Phys. Rev. C {\bf 48}, 1518 (1993). 
\bibitem{bulgac06} A. Bulgac, J.~E. Drut, and P.~Magierski, Phys. Rev. Lett. {\bf 96}, 090404 (2006).
\bibitem{stoicheva07} G. Stoitcheva, W.E. Ormand, D. Neuhauser, D.J. Dean, arXiv:0708.2945v1 (2007).
\bibitem{yu06} Z.~Yu and G. Baym, Phys. Rev. A {\bf 73}, 063601 (2006).
\bibitem{zoller00} P. T\"orm\"a and P. Zoller, Phys. Rev. Lett. {\bf 85}, 487 (2000).
\bibitem{leskinen08} M.~J.~Leskinen, V.~Apaja, J.~Kajala and P.~T\"orma, Phys. Rev. A {\bf 78}, 023602 (2008).
\bibitem{bulgac07} A. Bulgac, Phys. Rev. A {\bf 76}, 040502(R) (2007).
\bibitem{carlson05} J. Carlson and S. Reddy, Phys. Rev. Lett. {\bf 95}, 060401 (2005).
\bibitem{kinnunen04} J. Kinnunen, M. Rodriguez, and P. T\"orma, Science {\bf 305}, 1131 (2004).
\bibitem{stoof08} P.~Massignan, G.~M. Bruun, and H. T. C. Stoof, Phys. Rev. A {\bf 77}, 031601(R) (2008).
\bibitem{he05} Y.~He, Q.~Chen, and K.~Levin, Phys. Rev. A {\bf 72}, 011602(R) (2005).
\bibitem{chen05} Q.~Chen, J.~Stajic, and K.~Levin, Phys. Rev. Lett. {\bf 95}, 260405 (2005).
\bibitem{ormand94} W. E. Ormand, D. J. Dean, C. W. Johnson, G. H. Lang and S. E. Koonin, Phys Rev C {\bf 49}, 1422 (1994). 
Y. Alhassid, Rev. Mod. Phys. {\bf 72}, 895 (2000).
\end{thebibliography}
\end{document}